# Ice grains grow by dissolution, ripening and boundary migration

Henry Chan[1§], Mathew J. Cherukara[2§1], Badri Narayanan[1§], Chris Benmore[2], Stephen K. Gray[1, 3], Subramanian K.R.S. Sankaranarayanan[1, 3]

[1]Center for Nanoscale Materials, Argonne National Laboratory, Argonne, IL, 60439
[2]X-ray Science Division, Argonne National Laboratory, Argonne, IL, 60439
[3]Computation Institute, University of Chicago

**Abstract**

**Ice grain growth is ubiquitous, impacting naturally occurring phenomena such as glacier formation, global phenomena like climate change, and nanoscale processes like intracellular freezing. Grain size impacts the mechanical, thermal and optical properties of ice, yet an atomistic picture of ice grain formation has remained elusive. Despite the exponential growth in computing resources and the availability of a myriad of different theoretical water models, an accurate, yet computationally efficient molecular level description of mesoscopic grain growth remains a grand challenge. The underlying phase transitions and dynamical processes in deeply supercooled systems are often rendered inaccessible due to limitations imposed by system sizes and timescales, which is further compounded by their sluggish kinetics. Here, we introduce a machine-learned, bond-order-based potential model (BOP) that more accurately describes the anomalous behavior, as well as structural and thermodynamical properties of both liquid water and ice, with at least two orders of magnitude cheaper computational cost than existing atomistic water models. In a significant departure from conventional force-field fitting, we use a multilevel evolutionary strategy that trains the BOP model against not just energetics but

§ Equal contributions

**temperature dependent properties inferred from on-the-fly molecular dynamics simulations as well. We use the BOP model to probe the homogeneous ice nucleation and growth process by performing molecular dynamics on multi-million molecule systems for up to microsecond time scales. These massively parallel, long-time simulations naturally capture the competition between cubic, hexagonal, and stacking disordered ice phases during nucleation and early stages of growth leading to the formation of nanometer sized grains. Subsequently, we elucidate the hitherto elusive mechanism of grain coarsening of the crystallized mixed ice phases, which occur through grain dissolution and Ostwald ripening followed by grain boundary migration. This sequence of grain coarsening processes is unique to ice and has ramifications for a variety of processes from atmospheric cloud science, polar ice cap melting to modern day technologies like cryogenic food storage.**

The evolution of polycrystalline grains of ice formed from deeply supercooled water into larger crystallites is ubiquitous to diverse natural phenomena ranging from snowfalls, glacier formation, climate change to extraterrestrial objects like comets, as well as modern day technologies such as cryopreservation of food and biological samples. Ice nuclei when formed are nanoscopic[1–3] – critical sizes range from tens of molecules – whereas the grain sizes of most commonly observed forms of ice are in the millimeter to centimeter range[4–6]. As in most polycrystalline materials, grain size impacts its mechanical, physical, thermal and optical properties[7–11]. For instance, the grain size distribution in the polar ice sheets covering the Arctic Ocean dictates the ability of these sheets to act as a heat barriers (solar reflectors) and the subsequent heat release when they break up[12–14]. Likewise, the optical properties of snow vary intricately with grain size, particle shape and microscopic crystal surface roughness[14–16]. Similarly, the rheological properties of glaciers, ice sheets, polar ice caps, and icy planetary interiors are controlled in large by the grain-scale deformation of ice[17–23]. The dynamical processes at the evolving grain boundaries can combine the remarkable complexity of a multitude of different nucleating ice phases, their transport phenomena and microstructural evolution with the formidable subtleties of defect and solvation dynamics[24–31]. It is thus not surprising that the fundamental origins and sequence of steps from nucleation to ice grain formation and growth remain largely unknown.

An atomistic or molecular mechanism of the grain boundary formation, growth and consolidation of ice during the post-nucleation stage is most desirable but the corresponding length and timescales remain inaccessible to existing fully atomistic simulations[32]. Crystallization of supercooled water during the early stages occurs over several tens of nanometers and the subsequent grain growth extends over micron length scales. The typical timescales involved in the growth processes extend over tens of nanoseconds to several microsecond time scales and it is therefore extremely difficult to understand the molecular level phenomena associated with the ice grain growth that typically falls in the mesoscopic regime. While the move towards exascale computing will address length scale challenges, given that the clock speeds and bandwidths will not increase substantially, timescale challenges will remain critical (see Extended Data Fig. 1). It is therefore critical to have a water model that accurately captures the melting point, the densities of liquid water and ice, as well as other thermodynamic and transport properties at modest computational cost.

Over the past several decades, numerous atomistic[33–40] and coarse-grained[41–43] (CG) models have been developed in an attempt to describe the thermodynamic properties and dynamical behavior of water. These models differ in terms of the tradeoff between predictive power and computational cost/efficiency. The best performing atomistic model is TIP4P/2005[40]; it however under predicts the melting transition by 20 K and is far too computationally expensive for large-scale molecular dynamics (MD) studies of grain growth. CG models are a viable alternative with several orders of magnitude improvements in computational efficiency, allowing million-atom simulations to effortlessly reach microsecond time scales. Although such improvements in computational efficiency are needed for mechanistic understanding grain formation and growth, this gain comes at the expense of accuracy (see supplementary methods). The monoatomic water (mW) model[44], inspired by the similarity between water and other tetrahedral solids like Si and C, remains the best performing CG model; it predicts the correct melting point and thermodynamic properties such as enthalpy of fusion but does not quantitatively capture the density anomaly and over predicts the ice density in the 200-273 K range. Despite tremendous efforts, capturing the physical and thermodynamic properties of water notably the density anomaly, the melting transition and the relative density difference between ice/liquid water remains a grand challenge for all water models[45].

Here, we report on the development of a machine-learnt coarse-grained bond order potential (BOP) model that outperforms existing models in describing the structure, thermodynamic and transport properties of both ice and liquid water. We achieve this while simultaneously making significant improvements in computational efficiency: we are at least two-three orders of magnitude cheaper compared to the most accurate atomistic models (TIP4P and TIP5P). Motivated by the remarkable success of the mW model[44] noted above, our model also treats each water molecule as one bead, and with a potential form capable of describing tetrahedral solids (see Supplementary Information for the formalism and coarse graining details we employ). Describing the myriad and complex properties of water adequately with a potential model, even in coarse grained form, is a very challenging task. We therefore devise a multilevel evolutionary strategy that rigorously trains the BOP model against an extensive energetics and dynamical data set derived from the best available atomistic model (TIP4P/2005) supplemented by experimental data (see Supplementary Methods for details on the training data set). Note that this multilevel optimization procedure can be used to improve the predictive power of existing atomistic and coarse grained water models. Figure 1a outlines the training scheme used to optimize the BOP model. We start by generating an elaborate training dataset of TIP4P/2005 generated energetic and structural properties of both ice and liquid water. This departure from the sparse experimental data set used in most force field parameterization ensures an adequate representation of the diverse configurational space typical of ice and liquid water while amply sampling the energy landscape. We next use a combination of supervised machine-learning algorithm (*via* a genetic algorithm) to perform a global search followed by a local optimization (*via* the simplex method) to find the best set of the independent parameters for our BOP. This two-stage training scheme circumvents the problems encountered with the widely employed local minimization for force field fitting that overly rely on good starting guesses and often get trapped in local minima (Figure 1b). Our training strategy also marks an important paradigm shift from existing fitting procedures: one of the objectives involves minimizing properties calculated from nanosecond long trajectories derived from isobaric MD of ice and water using the BOP model *vs.* those from experiments at several temperatures in the 200-350 K range. Thus, numerous MD simulations are a key element of the iterative process employed to obtain the final BOP potential energy function.

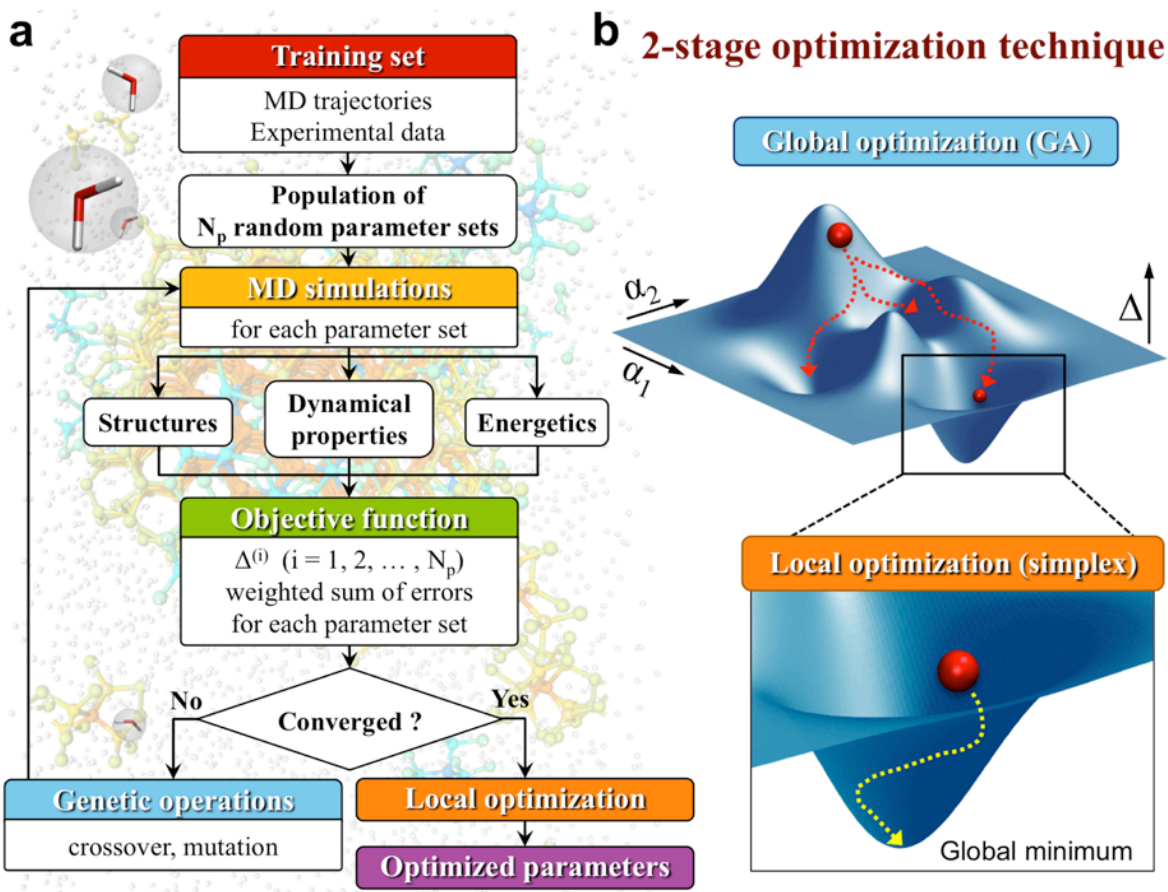

**Figure 1 | Illustration of the machine learning protocol employed to train the bond-order potential (BOP) for water**. (**a**) Workflow depicting the various stages of the force field parameterization. The novelty of this workflow is a direct fitting to dynamically-inferred properties through running and sampling MD simulations during the fitting process. (**b**) Schematic diagrams illustrating the concept of a 2-stage optimization technique used in our workflow for locating the global minimum of the objective landscape. The actual optimization spanned an 11-dimensional landscape but here we indicate just two dimensions.

Figure 2 compares the variation in the structural and dynamically-inferred properties including the anomalous behavior of water against available experimental data. BOP successfully captures the best-known thermodynamic anomaly *i.e.* the existence of density maximum at 4 C (Figure 2a). Quantitatively, the BOP model freezing/melting transition occurs at 273 +/- 1K and its predictions of the densities of ice (140-273 K) and water (243-373 K) are within 1.3% of their actual values. The existence and location of the temperature of maximum density (TMD) has been a challenge for most inter-atomic water potentials. TIP4P/2005 is the best atomistic model to depict TMD but it underestimates the melting point by 20 K. As shown in Extended Data Fig. 3, the BOP model is the most accurate theoretical framework that has been developed for water, outperforming previous models in almost all measurable aspects (See Supplementary Table 2). A comparison of BOP transport properties with experiments is shown in Figure 2b. The room temperature diffusivity of 3 x $10^{-5}$ cm$^2$/s predicted by the BOP model is in excellent agreement with experiment (2.3 x $10^{-5}$ cm$^2$/s). The BOP predicted diffusivities are in good agreement with experiment; the BOP model slightly overestimates the diffusivities in the supercooled range but outperforms other existing coarse-grained models (Extended data Fig. 3b). Figure 2c compares the oxygen-oxygen radial distribution function for ice I*h* at 77 K and (supercooled) liquid water at 254 K derived from X-ray scattering/neutron experiments with that from our model. The location and intensities of the peaks corresponding to first, second and third coordination shells are in excellent agreement. The BOP model captures the temperature dependent trend of these peaks (change in location and intensities), and our calculated number of water neighbors in the first solvation shell, integrated out to the predicted temperature independent isosbestic point (r = 3.25Å), is 4.6 which compares well with the 4.3-4.7 range for neutron/x-ray refined structures[46,47]. Also, the angular distribution function at 298 K agrees very well with TIP4P/2005 (Extended data Fig 4b). Figure 2d presents the heat capacities of liquid water and BOP model, with respect to their values at 309 K. Our model reproduces the thermodynamic anomaly indicated by the sharp increase in $C_p$ of supercooled water. The BOP model captures the associated liquid transformation to low-density amorphous phase and predicts a transformation temperature of 203 K, which is slightly lower compared to experimental value of 222 K (but is still better than existing water models such as TIP4P and mW). Overall, the predictions of the BOP model are either better or on par with the best available water models (Supporting Information Table 2).

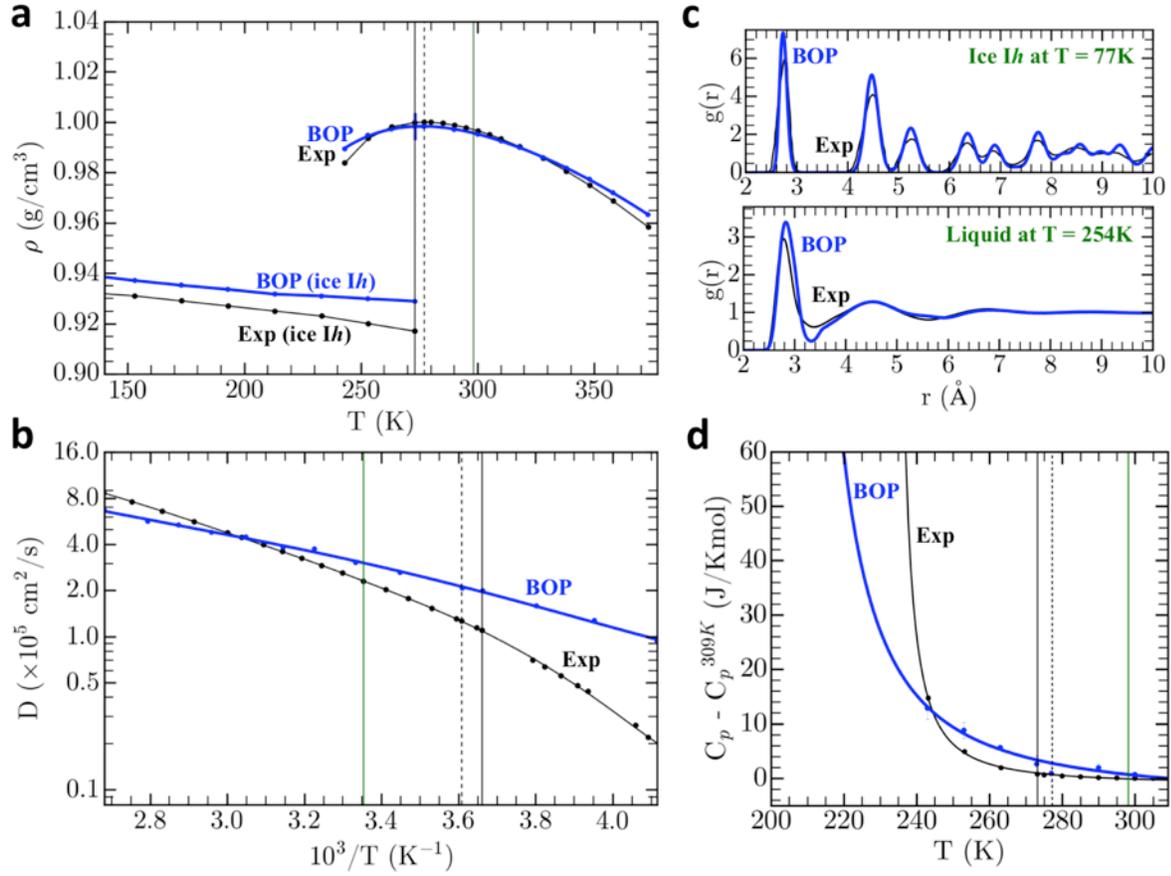

**Figure 2 | Comparison of the predictions of the BOP model with experiment (Exp).** In all plots, the experimental melting point (T=273 K), temperature of maximum density (T=277K), and room temperature (T=298 K) are marked by a vertical solid black line, dotted black line, and solid green line, respectively. (**a**) The BOP model is able to accurately reproduce the density anomaly of water within 1.3% error as shown by the comparison with experimental densities[48] of ice and liquid water at pressure P = 1 bar. A cross on the liquid density marks the predicted melting point (T=273±1K). (**b**) The BOP model predicts the experimental diffusion coefficients of water[49,50] over a wide temperature range. The y-axis is shown in log scale. (**c**) The BOP model reproduces the experimental radial distribution functions of ice at T=77 K[51] and liquid water at T=254 K[46]. (**d**) The BOP model captures the experimental heat capacity of water[52] relative to the value at T=309 K. The fitting equation $f(T, a, b, c) = a\left(\frac{T}{b} - 1\right)^{-1.5} + c$ is used to extrapolate the data into the supercooled region (T<245K). Both curves are offset by the value at T=309K (37.55 for BOP and 75.29 for Exp) for easy comparison. The curve fitting parameters for BOP are a=1.43 J/K-mol, b=203.72 K, c=33.69 J/K-mol) and for Exp are a=0.15 J/K-mol, b=232.80 K, c=-1.1 J/K-mol).

We perform MD on multi-million water molecules using this newly developed BOP to understand the molecular level sequence of steps starting from homogeneous nucleation of supercooled water leading up to the formation and growth of grains of ice. Figure 3 summarizes the initial stages of formation of polycrystalline ice for one such trajectory when water is slowly cooled from 275 K to 210.5 K over 258 ns (cooling rate ~ 2.5 x $10^8$ K/s). Following the appearance of the first stable nuclei at ~210.5

K, the temperature was held at 210.5 K for a further 110 ns to study the nucleation and growth processes in this homogeneously nucleated water. Figure 3a shows the potential energy variation as a function of time during the cooling phase and constant temperature phase. We identify four distinct stages during the freezing process (a) a long quiescent time period of ~250 nanoseconds before the first nucleation events, (b) a period of slow transformation with a limited number of nuclei (8 at t = 270 ns), (c) accelerated transformation driven by growth of a greater number of nuclei (~80 at 300 ns), and (d) completion of grain growth to form a polycrystalline box of ice. Figure 3b shows the corresponding snapshots during the initial quiescent period when the system explores the relatively flat energy landscape before entering the nucleation and growth period. This molecular level illustration is consistent with classical nucleation theory; the quiescent period is marked by pronounced fluctuation of many subcritical nuclei which rapidly form, break and reform in the supercooled liquid as shown in Figure 3d. The post-quiescent period shown by MD snapshots in Figure 3c is marked by formation of multiple stable nuclei which grow slowly followed by a rapid growth phase when the grains begin to percolate through the entire three-dimensional space. The completion of the growth phase is characterized by the formation of a polycrystalline ice with the nanoscopic grains separated by boundaries comprised of amorphous ice. A local structure analysis of the growing structure reveals that the grains are comprised of stacking disordered ice (I$SD$) *i.e.* randomly mixed alternating sheets of hexagonal and cubic ice (Figure 3c). The evolution, extent and relevance of stacking disorder in polycrystalline ice has been a matter of much debate. Our MD simulations unambiguously capture the competition between the cubic (I$c$) and hexagonal (I$h$) phases leading up to the formation of I$SD$ at atmospheric conditions. Figure 3e shows that the evolving ice structure becomes increasingly rich in I$c$ compared to the more stable I$h$ phase with the fraction of cubic to hexagonal to be ~ 1.6 at the end of t = 350 ns. The observed preference for cubic ice formation is consistent with multiple experimental results in the past including a recent X-ray diffraction study[53] and coarse-grained simulations[54] as well as atomistic simulations using forward-flux sampling technique[55].

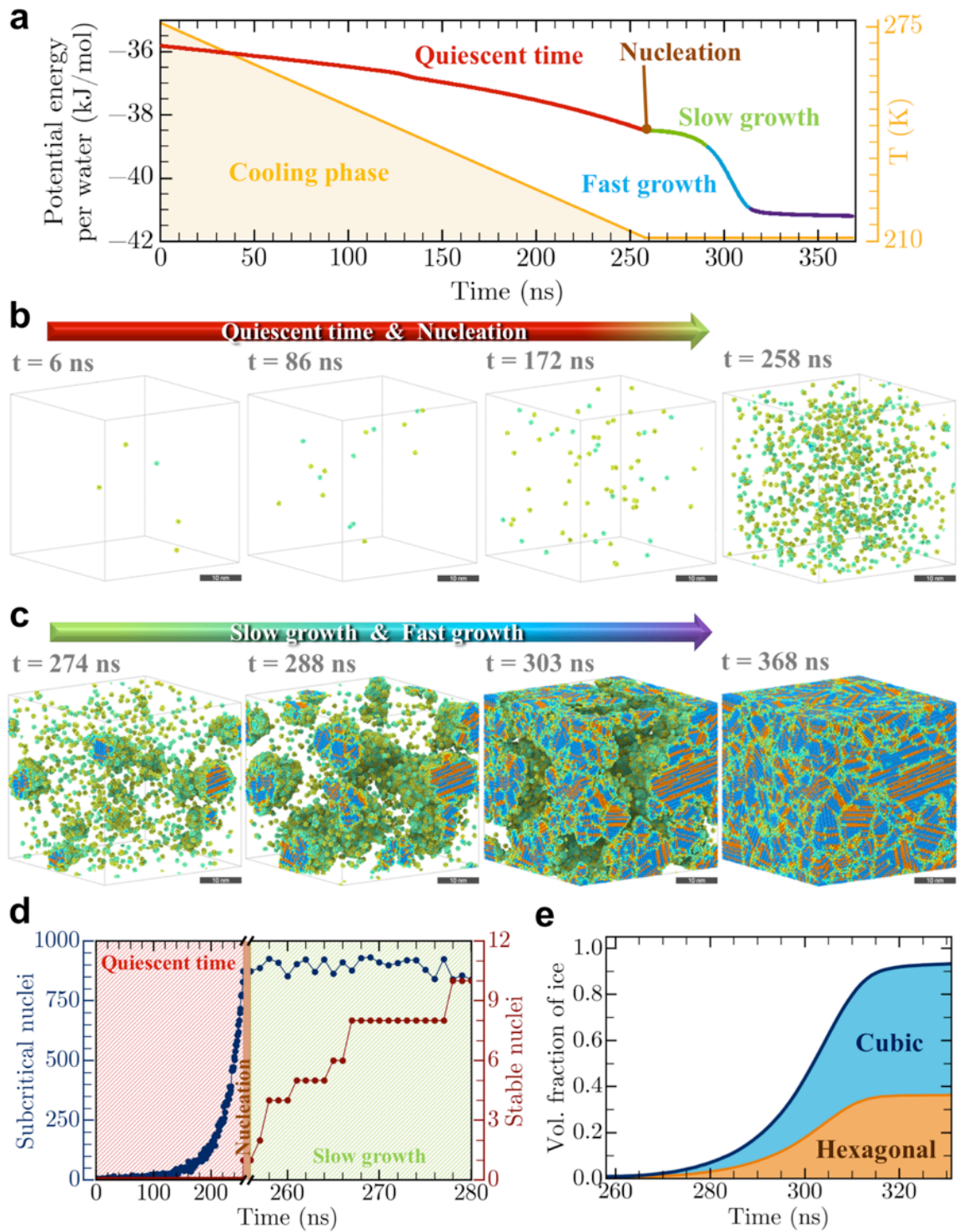

**Figure 3 | System dynamics and evolution of structural motifs during the cooling phase from homogeneous nucleation leading up to the grain boundary formation and grain growth.** (**a**) The total potential energy variation of the 2 million-water molecule system during the cooling phase from 275 to

210.5 K and at longer times when the system temperature is kept constant at 210.5 K. We identify four distinct stages (i) Initial quiescent time shown by the red line when no nucleation event occurs (ii) the nucleation followed by an initial slow transformation shown by the slow energy decreasing period in green (iii) a fast transformation phase of the grains shown by the rapid decrease in potential energy in blue (iv) a plateauing of potential energy shown in purple marks the completion of the phase transformation. (**b**) The snapshots showing the subcritical water nuclei during the long quiescent phase leading up to the nucleation. The first nucleation event for the 2 million-water system occurs at t = 258 ns. Liquid water molecules are not shown for clarity. (**c**) MD simulation snapshots showing the various stages of grain growth and grain boundary during the post-nucleation stage. Blue, brown and green spheres represent cubic, hexagonal and amorphous ice, respectively. Liquid water is omitted for clarity. (**d**) The temporal evolution of the number of subcritical water nuclei (size < 100 molecules) from the quiescent period and the initial appearance of stable nuclei during the post-nucleation stage. (**e**) The corresponding temporal evolution of the fraction of cubic and hexagonal ice.

The microstructure obtained at the conclusion of the cooling and constant temperature simulation (Figure 3c) is extremely fine grained (average grain size ~15,000 water molecules). This fine microstructure is expected to anneal over long times (microsecond to seconds) to naturally observed grain sizes that range from micro-meters to millimeters in size. An atomistic picture of the post-nucleation coarsening of grains remains largely elusive. To study the molecular rearrangement processes driving the transformation to large grain sizes, we anneal at 260 K (a typical temperature attained by glaciers due to seasonal variations[56]) the nanocrystalline sample that was obtained from quenching and holding at 210 K. We observe two mechanisms that drive the annealing of ice crystallites. Initially, we observe concurrent dissolution of small grains (grain size < 2000 water molecules) and growth of large ones (grain size > 11,000 water molecules) analogous to the Ostwald ripening process in solution. Within ~10 ns of annealing, *i.e.* t = 368-378 ns (red curve Figure 4a), these small grains, owing to their low stability (high surface to volume ratio) melt away; the water molecules from this melt subsequently impinge, and contribute to the growth of nearby larger grains (Figure 4b). This interplay between melting and grain growth leads to observed maxima and minima at t = 371 ns in the fraction of amorphous and crystalline phases, respectively. The larger grains continue to grow into the space vacated by the dissolved grains, and quickly occupy the entire volume of the box (t = 378-430 ns). This marks the beginning of a second stage of slower growth where grain coarsening continues through boundary migration (t = 430 ns and beyond) until only 2 grains remain in the box (Supplementary Fig. 1). Figure 4c and Video 4 shows the grain boundary migration mechanism. The zoomed in images/video shows the consumption of a grain by its neighbors over a period of ~30 ns. Bar graphs below each frame show the grain size distribution over the entire simulation cell and reveal the progression to larger grain sizes.

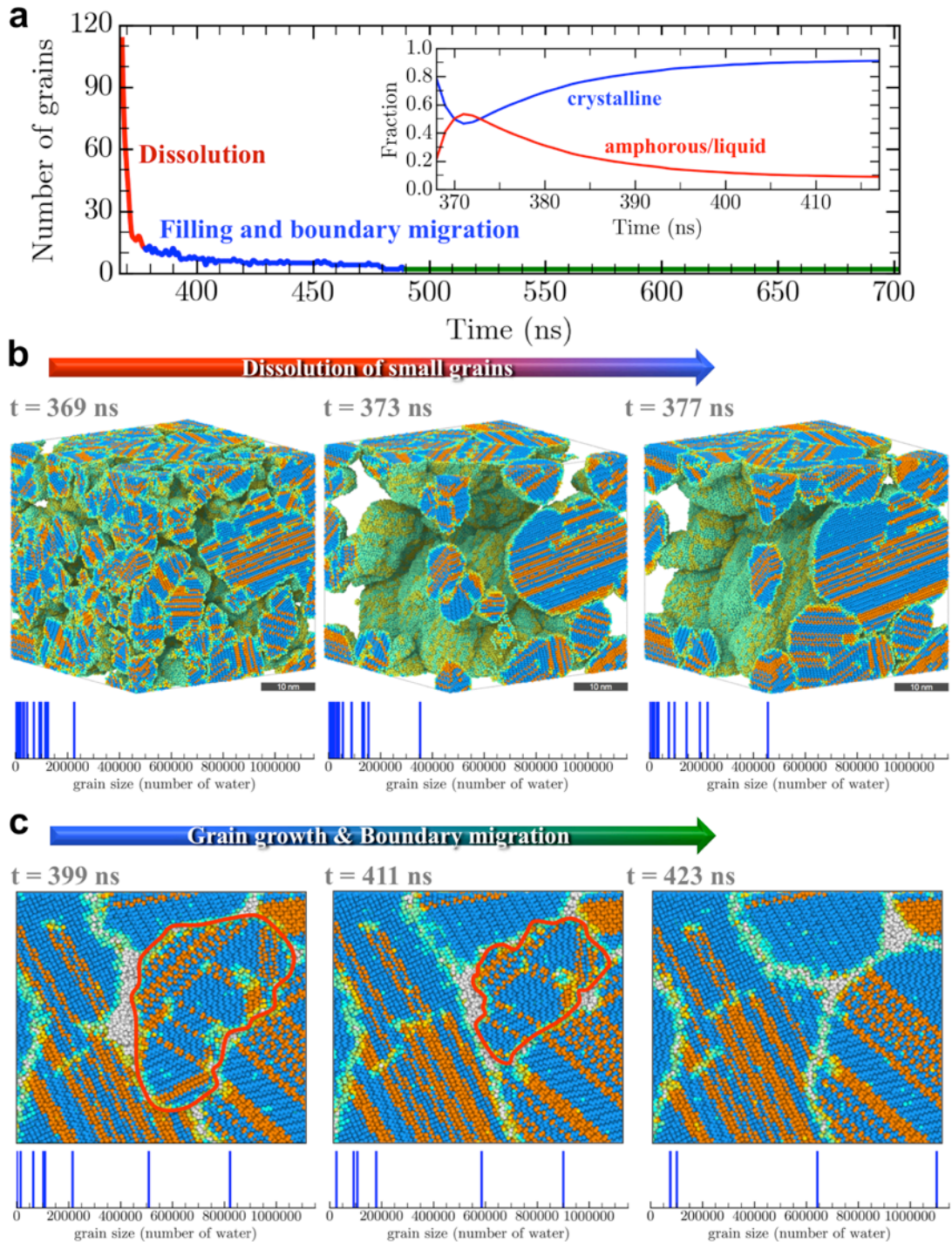

**Figure 4 | Mechanism of formation of grain boundaries and grain growth in ice nucleated from supercooled water.** (a) Temporal evolution of the number of grains in polycrystalline ice during annealing at 260 K. We identify two distinct stages: (i) an initial Ostwald ripening stage where we observe a sharp drop in the number of grains at t < 10 ns due to rapid dissolution of smaller sized grains into amorphous ice as indicated by the red line, and (ii) a slower grain consolidation phase which occurs due to grain boundary migration indicated by the blue line followed by a stable phase indicated by green line when the number of grains remains unchanged. Inset shows the corresponding temporal evolution of

the grain size. (b) The fractional change in the amorphous and crystalline ice during the grain dissolution phase. The maxima and minima in the amorphous and crystalline population, respectively result from the relative kinetics of grain dissolution and grain growth. (c) Snapshots from the multi-million MD trajectory show the rapid dissolution of smaller sized grains into amorphous ice whereas the larger energetically stable grains survive and grow by consuming the amorphous ice (The amorphous ice is not shown for clarity but readers may examine Video 3 that shows the dissolution of small grains to amorphous ice (represented by white spheres)). The panel below depicts the bar graph of grain size distribution at the time instant corresponding to the snapshot. Initially, at t = 369 ns, there is a large population of smaller sized grains. With time, the population of smaller sized grains progressively decreases whereas that of larger grains increases. (d) Zoomed-in view of snapshots from the multi-million MD trajectories showing the grain growth via the classic grain boundary migration. The larger grains are shown to grow at the expense of smaller grains. The grain size distribution below shows the associated progression towards larger sized grains as a result of grain boundary migration.

In summary, we introduced a computationally efficient coarse-grained potential model to accurately identify the molecular mechanisms underlying grain formation and growth in ice starting from supercooled liquid water. We demonstrated that grain dissolution followed by Ostwald ripening, a well-known mechanism for precipitation in solutions, acts as a precursor for a second, slower coarsening mechanism, namely grain boundary migration in ice. Both these mechanisms are crucial for the realization of millimeter size grains observed in nature. The grain size distribution and its variation impact a broad range of natural phenomena including glacier melting/drifting, cloud formation and other atmospheric processes influencing global warming. For instance, climate change models have been shown to be extremely sensitive to number and size distribution of ice nuclei in the upper atmosphere[57]. Likewise, the evolution and distribution of the average grain size in the deep ice cores impacts its strain rate dependence and thereby the erosion of polar ice[58]. We also anticipate that these grain growth mechanisms are applicable for a broad class of low melting solids such gallium, indium, etc. as well as supercooled aqueous solutions.

**References:**

1. Bigg, E. K. & Hopwood, S. C. Ice Nuclei in the Antarctic. *J. Atmospheric Sci.* **20,** 185–188 (1963).

2. Majewski, J. *et al.* Toward a Determination of the Critical Size of Ice Nuclei. A Demonstration by Grazing Incidence X-ray Diffraction of Epitaxial Growth of Ice under the C31H63OH Alcohol Monolayer. *J. Phys. Chem.* **98,** 4087–4093 (1994).

3. Liu, J., Nicholson, C. E. & Cooper, S. J. Direct Measurement of Critical Nucleus Size in Confined Volumes. *Langmuir* **23,** 7286–7292 (2007).

4. Gow, A. J. *et al.* Physical and structural properties of the Greenland Ice Sheet Project 2 ice core: A review. *J. Geophys. Res. Oceans* **102,** 26559–26575 (1997).

5. Worby, A. P., Massom, R. A., Allison, I., Lytle, V. I. & Heil, P. in *Antarctic Sea Ice: Physical Processes, Interactions and Variability* (ed. Jeffries,  rtin O.) 41–67 (American Geophysical Union, 1998).

6. Faria, S. H., Weikusat, I. & Azuma, N. The microstructure of polar ice. Part I: Highlights from ice core research. *J. Struct. Geol.* **61,** 2–20 (2014).

7. Petrovic, J. J. Review Mechanical properties of ice and snow. *J. Mater. Sci.* **38,** 1–6

8. Schulson, E. M. & Duval, P. *Creep and Fracture of Ice*. (Cambridge University Press, 2009).

9. Weeks, W. F. & Lee, O. S. Observations on the Physical Properties of Sea-Ice at Hopedale, Labrador. *ARCTIC* **11,** 134–155 (1958).

10.     Yen, Y.-C. *Review of Thermal Properties of Snow, Ice and Sea Ice,*. (1981).

11.     Perovich, D. K., Roesler, C. S. & Pegau, W. S. Variability in Arctic sea ice optical properties. *J. Geophys. Res. Oceans* **103,** 1193–1208 (1998).

12. Doran, P. T. *et al.* Climate forcing and thermal feedback of residual lake-ice covers in the high Arctic. *Limnol. Oceanogr.* **41,** 839–848 (1996).

13. Ushio, S. Factors affecting fast-ice break-up frequency in Lützow-Holm Bay, Antarctica. *Ann. Glaciol.* **44,** 177–182 (2006).

14. Grenfell, T. C., Warren, S. G. & Mullen, P. C. Reflection of solar radiation by the Antarctic snow surface at ultraviolet, visible, and near-infrared wavelengths. *J. Geophys. Res. Atmospheres* **99,** 18669–18684 (1994).

15. Warren, S. G. Optical properties of snow. *Rev. Geophys.* **20,** 67–89 (1982).

16. Kokhanovsky, A. A., Aoki, T., Hachikubo, A., Hori, M. & Zege, E. P. Reflective properties of natural snow: approximate asymptotic theory versus in situ measurements. *IEEE Trans. Geosci. Remote Sens.* **43,** 1529–1535 (2005).

17. Boulton, G. S. & Hindmarsh, R. C. A. Sediment deformation beneath glaciers: Rheology and geological consequences. *J. Geophys. Res. Solid Earth* **92,** 9059–9082 (1987).

18. Budd, W. F. & Jacka, T. H. A review of ice rheology for ice sheet modelling. *Cold Reg. Sci. Technol.* **16,** 107–144 (1989).

19. Gow, A. J. & Williamson, T. Rheological implications of the internal structure and crystal fabrics of the West Antarctic ice sheet as revealed by deep core drilling at Byrd Station. *Geol. Soc. Am. Bull.* **87,** 1665–1677 (1976).

20. Fisher, D. A. & Koerner, R. M. On the Special Rheological Properties of Ancient Microparticle-Laden Northern Hemisphere Ice as Derived from Bore-Hole and Core Measurements. *J. Glaciol.* **32,** 501–510 (1986).

21. Mangold, N., Allemand, P., Duval, P., Geraud, Y. & Thomas, P. Experimental and theoretical deformation of ice–rock mixtures: implications on rheology and ice content of Martian permafrost. *Planet. Space Sci.* **50,** 385–401 (2002).

22. Durham, W. B., Kirby, S. H. & Stern, L. A. Effects of dispersed particulates on the rheology of water ice at planetary conditions. *J. Geophys. Res. Planets* **97,** 20883–20897 (1992).

23. Durham, W. B. & Stern, L. A. Rheological Properties of Water Ice—Applications to Satellites of the Outer Planets. *Annu. Rev. Earth Planet. Sci.* **29,** 295–330 (2001).

24. Durand, G., Graner, F. & Weiss, J. Deformation of grain boundaries in polar ice. *Europhys. Lett. EPL* **67,** 1038–1044 (2004).

25. Montagnat, M. & Duval, P. Rate controlling processes in the creep of polar ice, influence of grain boundary migration associated with recrystallization. *Earth Planet. Sci. Lett.* **183,** 179–186 (2000).

26. Blackford, J. R. Sintering and microstructure of ice: a review. *J. Phys. Appl. Phys.* **40,** R355 (2007).

27. Montagnat, M., Duval, P., Bastie, P., Hamelin, B. & Lipenkov, V. Y. Lattice distortion in ice crystals from the Vostok core (Antarctica) revealed by hard X-ray diffraction; implication in the deformation of ice at low stresses. *Earth Planet. Sci. Lett.* **214,** 369–378 (2003).

28. Montagnat, M. & Duval, P. The viscoplastic behaviour of ice in polar ice sheets: experimental results and modelling. *Comptes Rendus Phys.* **5,** 699–708 (2004).

29. Thomas, D. N. & Dieckmann, G. S. *Sea Ice: An Introduction to its Physics, Chemistry, Biology and Geology*. (John Wiley & Sons, 2008).

30.  Li, W. L., Lu, K. & Walz, J. Y. Freeze casting of porous materials: review of critical factors in microstructure evolution. *Int. Mater. Rev.* **57,** 37–60 (2012).

31.  Weikusat, I., Kipfstuhl, S., Faria, S. H., Azuma, N. & Miyamoto, A. Subgrain boundaries and related microstructural features in EDML (Antarctica) deep ice core. *J. Glaciol.* **55,** 461–472 (2009).

32.  Sosso, G. C. *et al.* Crystal Nucleation in Liquids: Open Questions and Future Challenges in Molecular Dynamics Simulations. *Chem. Rev.* **116,** 7078–7116 (2016).

33.  Jorgensen, W. L., Chandrasekhar, J., Madura, J. D., Impey, R. W. & Klein, M. L. Comparison of simple potential functions for simulating liquid water. *J. Chem. Phys.* **79,** 926–935 (1983).

34.  Berendsen, H. J. C., Grigera, J. R. & Straatsma, T. P. The missing term in effective pair potentials. *J. Phys. Chem.* **91,** 6269–6271 (1987).

35.  Jorgensen, W. L. & Madura, J. D. Temperature and size dependence for Monte Carlo simulations of TIP4P water. *Mol. Phys.* **56,** 1381–1392 (1985).

36.  Mahoney, M. W. & Jorgensen, W. L. A five-site model for liquid water and the reproduction of the density anomaly by rigid, nonpolarizable potential functions. *J. Chem. Phys.* **112,** 8910–8922 (2000).

37.  Nada, H. & Eerden, J. P. J. M. van der. An intermolecular potential model for the simulation of ice and water near the melting point: A six-site model of H2O. *J. Chem. Phys.* **118,** 7401–7413 (2003).

38.  Horn, H. W. *et al.* Development of an improved four-site water model for biomolecular simulations: TIP4P-Ew. *J. Chem. Phys.* **120,** 9665–9678 (2004).

39. Abascal, J. L. F., Sanz, E., Fernández, R. G. & Vega, C. A potential model for the study of ices and amorphous water: TIP4P/Ice. *J. Chem. Phys.* **122,** 234511 (2005).

40. Abascal, J. L. F. & Vega, C. A general purpose model for the condensed phases of water: TIP4P/2005. *J. Chem. Phys.* **123,** 234505 (2005).

41. Hadley, K. R. & McCabe, C. Coarse-Grained Molecular Models of Water: A Review. *Mol. Simul.* **38,** 671–681 (2012).

42. Darré, L., Machado, M. R. & Pantano, S. Coarse-grained models of water. *Wiley Interdiscip. Rev. Comput. Mol. Sci.* **2,** 921–930 (2012).

43. Orsi, M. Comparative assessment of the ELBA coarse-grained model for water. *Mol. Phys.* **112,** 1566–1576 (2014).

44. Molinero, V. & Moore, E. B. Water Modeled As an Intermediate Element between Carbon and Silicon. *J. Phys. Chem. B* **113,** 4008–4016 (2009).

45. Agarwal, M., Alam, M. P. & Chakravarty, C. Thermodynamic, Diffusional, and Structural Anomalies in Rigid-Body Water Models. *J. Phys. Chem. B* **115,** 6935–6945 (2011).

46. Skinner, L. B., Benmore, C. J., Neuefeind, J. C. & Parise, J. B. The structure of water around the compressibility minimum. *J. Chem. Phys.* **141,** 214507 (2014).

47. Soper, A. K. The Radial Distribution Functions of Water as Derived from Radiation Total Scattering Experiments: Is There Anything We Can Say for Sure? *Int. Sch. Res. Not.* **2013,** e279463 (2013).

48. *CRC handbook of chemistry and physics: a ready-reference book of chemical and physical data*. (CRC Press, 2004).

49. Gillen, K. T., Douglass, D. C. & Hoch, M. J. R. Self-Diffusion in Liquid Water to −31°C. *J. Chem. Phys.* **57,** 5117–5119 (1972).

50. Holz, M., Heil, S. R. & Sacco, A. Temperature-dependent self-diffusion coefficients of water and six selected molecular liquids for calibration in accurate 1H NMR PFG measurements. *Phys. Chem. Chem. Phys.* **2,** 4740–4742 (2000).

51. Narten, A. H., Venkatesh, C. G. & Rice, S. A. Diffraction pattern and structure of amorphous solid water at 10 and 77 °K. *J. Chem. Phys.* **64,** 1106–1121 (1976).

52. Chickos, J. S. & Jr, W. E. A. Enthalpies of Sublimation of Organic and Organometallic Compounds. 1910–2001. *J. Phys. Chem. Ref. Data* **31,** 537–698 (2002).

53. Malkin, T. L., Murray, B. J., Brukhno, A. V., Anwar, J. & Salzmann, C. G. Structure of ice crystallized from supercooled water. *Proc. Natl. Acad. Sci.* **109,** 1041–1045 (2012).

54. Moore, E. B. & Molinero, V. Is it cubic? Ice crystallization from deeply supercooled water. *Phys. Chem. Chem. Phys.* **13,** 20008–20016 (2011).

55. Haji-Akbari, A. & Debenedetti, P. G. Direct calculation of ice homogeneous nucleation rate for a molecular model of water. *Proc. Natl. Acad. Sci.* **112,** 10582–10588 (2015).

56. Perovich, D. K. & Elder, B. C. Temporal evolution of Arctic sea-ice temperature. *Ann. Glaciol.* **33,** 207–211 (2001).

57. DeMott, P. J. *et al.* Predicting global atmospheric ice nuclei distributions and their impacts on climate. *Proc. Natl. Acad. Sci.* **107,** 11217–11222 (2010).

58. Wellner, J. S., Lowe, A. L., Shipp, S. S. & Anderson, J. B. Distribution of glacial geomorphic features on the Antarctic continental shelf and correlation with substrate: implications for ice behavior. *J. Glaciol.* **47,** 397–411 (2001).

59. Vega, C., Abascal, J. L. F., Conde, M. M. & Aragones, J. L. What ice can teach us about water interactions: a critical comparison of the performance of different water models. *Faraday Discuss* **141,** 251–276 (2009).

60. Vega, C. & Abascal, J. L. F. Relation between the melting temperature and the temperature of maximum density for the most common models of water. *J. Chem. Phys.* **123,** 144504 (2005).

61. Lee, S. H. Temperature Dependence on Structure and Self-Diffusion of Water: A Molecular Dynamics Simulation Study using SPC/E Model. *Bull. Korean Chem. Soc.* **34,** 3800–3804 (2013).

62. Pi, H. L. *et al.* Anomalies in water as obtained from computer simulations of the TIP4P/2005 model: density maxima, and density, isothermal compressibility and heat capacity minima. *Mol. Phys.* **107,** 365–374 (2009).

63. Ester, M., Kriegel, H.-P., Sander, J. & Xu, X. A density-based algorithm for discovering clusters in large spatial databases with noise. in 226–231 (AAAI Press, 1996).

**Acknowledgements**

The authors thank Maria Chan, Alper Kinaci, Kiran Sasikumar, Al Wagner and Ross Harder for useful discussions. Use of the Center for Nanoscale Materials was supported by the U. S. Department of Energy, Office of Science, Office of Basic Energy Sciences, under Contract No. DE-AC02-06CH11357. This research also used resources of the Argonne Leadership Computing Facility at Argonne National Laboratory, which is supported by the Office of Science of the U.S. Department of Energy under contract DE-AC02-06CH11357. This research used resources of the National Energy Research Scientific Computing Center, a DOE Office of Science User Facility supported by the Office of Science of the U.S. Department of Energy under Contract No. DE-AC02-05CH11231. We also acknowledge the Carbon, Fusion and LCRC computing facilities at Argonne.

**Author Contributions**

HC, MJC and BN contributed equally. MJC, BN and SKRS conceived and designed the project. HC, BN, MJC, and SKRS developed the machine learning framework and bond order potential model for water with input from SKG. MJC performed the large scale simulations of grain formation and growth. HC, BN and MJC developed the feature detection algorithm for 3D analysis of grain size and distribution. SKRS supervised the overall project. All the authors including CJB and SKG performed the data analysis and contributed to the preparation of the manuscript.

**Methods:**

A machine learnt coarse grained bond order potential model for water is developed. Training data included an extensive data set of configurations, energies and MD trajectories of ice and liquid water at various temperatures derived from atomistic simulations using TIP4P. Experimental data set of melting point and other thermodynamic properties such as density, specific heat as well as their temperature dependence were also used in the training data set.

Details of the mapping of coarse grained model and the training procedure employed are provided in the supporting information. We performed massively parallel, long-time molecular dynamics simulations of super-cooled water using the BOP model to study the sequence of steps from nucleation to grain formation and coarsening. The simulation set up is as follows: we started with a simulation cell containing 2,048,000 million molecules of water at 275 K. MD simulations with a larger sized cell containing ~ 8 million water molecules were also performed. In both the cases, the structure was minimized and equilibrated for 100 ps at 275 K in an NPT ensemble. Subsequently, the liquid water was cooled down to 200 K over 300 ns (0.25 K/ns) under iso-baric conditions. We observed the first stable nuclei at ~211 K. Consequently, we stopped the cooling process at 210.5 K and held the super-cooled liquid at 210.5 K also under iso-baric conditions for 75 ns. Following the nucleation and growth of nanocrystalline ice grains at this temperature, we annealed the structure obtained at 260 K also under isobaric conditions for up to a microsecond to study the temporal evolution of grains. Annealing runs were also performed at 270 K for the sake of comparison. A feature detection algorithm based on an unsupervised machine learning technique was developed to perform the analysis of average grain sizes and grain size distribution. Details of the methods are provided in the Supplementary Information.

**Author information** Correspondence and requests for materials should be addressed to SKRS (skrssank@anl.gov)

# Extended Data

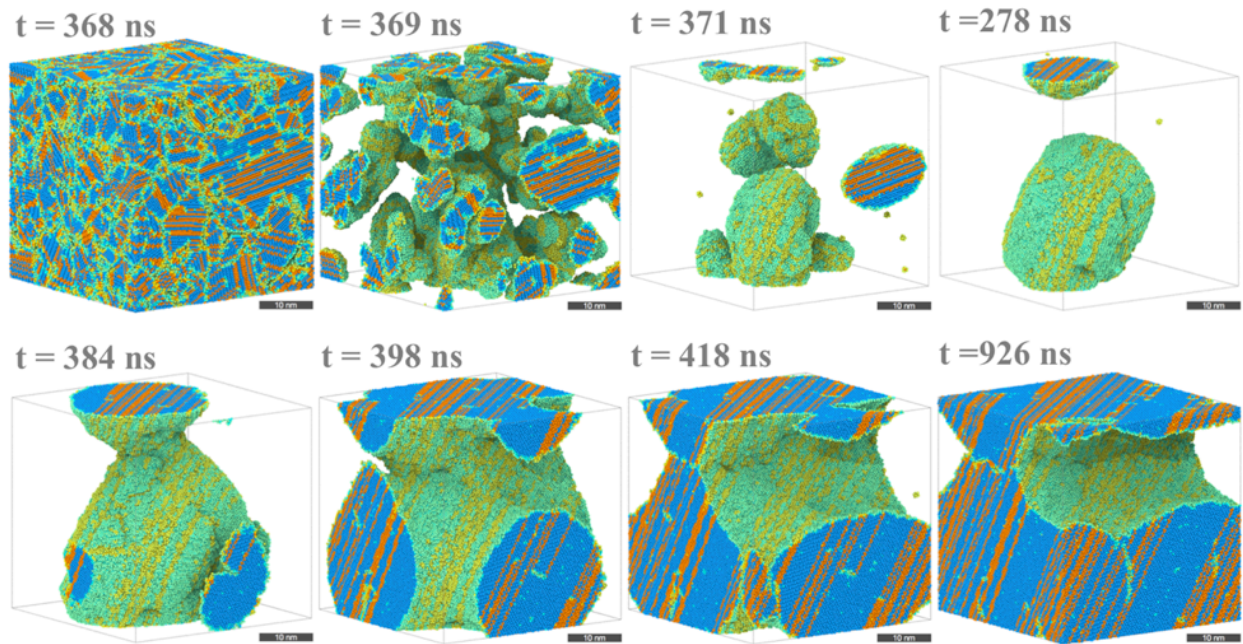

**Extended Data Figure 6** | Temporal evolution of an ice microstructure annealed at 270 K.